\documentstyle[12pt]{article}
\begin{document}
\title{Theory of hybrid systems. I. The operator formulation of
classical mechanics and semiclassical limit}
\author{Slobodan Prvanovi\'c and Zvonko Mari\'c \\ {\it Institute of
Physics, P.O. Box 57, 11080 Belgrade,} \\ {\it Yugoslavia}}
\date{}
\maketitle
\begin{abstract}
The algebra of polynomials in operators that represent generalized
coordinate and momentum and depend on the Planck constant is
defined. The Planck constant is treated as the parameter taking
values between zero and some nonvanishing $h_0$. For the second of
these two extreme values, introduced operatorial algebra becomes
equivalent to the algebra of observables of quantum mechanical
system defined in the standard manner by operators in the Hilbert
space. For the vanishing Planck constant, the generalized algebra
gives the operator formulation of classical mechanics since it is
equivalent to the algebra of variables of classical mechanical
system defined, as usually, by functions over the phase space. In
this way, the semiclassical limit of kinematical part of quantum
mechanics is established through the generalized operatorial
framework. 
\end{abstract}

\section{Introduction}

The hybrid systems are those consisting of two distinguished
subsystems. We are interested in situation when one of these subsystems
is quantum and the other one classical mechanical, where, of course, by
quantum mechanical system we mean such system whose behavior is
properly described in terms of quantum mechanics. Similarly, concepts
and relations of classical mechanics offer the complete formal
description of the second subsystem. On the other side, we believe that
``hybrid system'' is not bad terminological solution for a physical
system whose classical and quantum variants are simultaneously
considered.

Theory of hybrid systems should incorporate classical and quantum
mechanics on equal footing, together with transitions from one to
the other. The correspondence principle, or the quantization and
dequantization procedure, should be the essential characteristic
of this theory. In other words, it should offer mathematical
framework that supports simultaneous meaningful formulation of
both mechanics. Within it, classical and quantum mechanics should
appear as the same regarding the structural characteristics. In
particular, they should have one and the same algebraic product
and one and the same Lie bracket and, if one of them is formulated
in terms of differential geometry, then the other should be, too.
Moreover, fundamental entities of classical and quantum mechanics
should be interrelated in unambiguous and correct way.

Theory of hybrid systems should give the proper description of
interaction between quantum and classical systems. By proper we mean
that the collapse of state of quantum system, which can happen when
this system interacts with the classical one, {\it e.g.}, in the
process of measurement, should be explained/described without
going into circular argumentation.

With such theory one certainly would not solve all problems of
classical and quantum mechanics. This, after all, is not its
purpose. By exploiting the juxtaposition of classical and quantum
mechanics, this theory should give better insight in the two
mechanics and their relations. Eventually, this will be enough for
finding answers to some important questions.

After this rather general statements, let us give an outlook of
the program and strategy of our approach to the listed requests.
In the first paper of a series addressing hybrid systems we shall
propose new framework that offers simultaneous representations of
classical and quantum mechanics instead of using the framework of
functions over the phase space, see Argawal and Wolf (1970); Guz
(1984); Jordan and Sudarshan (1961); Mehta (1964); Misra and Shankara
(1968); Riccia and Wiener (1966); Shirokov (1976) and Uhlhorn (1956).
More concretely, we shall define the general algebra of operators that
depend on the Planck constant. We shall take this constant as the
parameter which takes value between some nonvanishing $h_0$ and zero.
For $h_0$ the algebra will reduce in one which is isomorphic to the
algebra of quantum mechanical system defined in standard fashion.
The other extreme value will be attributed to the classical
mechanics. That is, for the vanishing Planck constant, the
generalized operator algebra will reduce to the algebra that is in
one-to-one correspondence with the well-known algebra of variables
of classical mechanical system. Looked from another angle, this
will mean that there is an operatorial representation of classical
mechanics, as well as of quantum mechanics, and that semiclassical
limit (dequantization) is established on the kinematical level
through the general algebra. The dynamics will be just mentioned
in this paper since it asks for longer exposition.

The Lie algebraical structure of classical and quantum mechanics
is important for physics, see Emch (1972). The Lie bracket, or
product, in quantum mechanics is closely connected to the ordering
rule. In order to complete the correspondence principle, {\it
i.e.}, to include quantization, one has to analyze this topics as
well. This we shall do in the second article of this series by
defining the ordering rule - symmetrized product, of quantum
mechanics. This product will be the one according to which the
algebras in both classical and quantum mechanics should be formed.
In the classical mechanics, where one deals with the commutative
algebra, one can apply without problems any ordering rule, so one
can use symmetrized product as the algebraical one. Without going
into details, let us mention that by using the symmetrized product
we are going to propose, one can define the symmetrized Poisson
bracket which can substitute the commutator in quantum mechanics
becoming its Lie bracket. This bracket can be used in classical
mechanics as well (the explanation is as for the algebraic
product). Having one and the same algebraical product and one and
the same Lie bracket for both mechanics, we shall be able to
propose, we believe, unambiguous - obstruction free, quantization
prescription. (The ordering rule and obstructions to quantization
were previously discussed in Arens and Babit (1965); Chernoff (1995);
Cohen (1966); Gotay (1980) and (1996) and Kerner and Sutcliffe (1970).)
Moreover, the dynamical equations of both classical and quantum
mechanics will appear to be one and the same - the operatorial version
of the Liouville equation.

In the third article we shall unify the results of the first two.
Namely, we shall start with the general operators of coordinate
and momentum which depend on the Planck constant, then we shall
introduce one algebraical product and one Lie algebraical product
and, after considering the above mentioned extreme values of
Planck constant, we shall establish the semiclassical limit of
quantum mechanics for complete algebraical and Lie algebraical
settings. Then, because observables and variables are functions of
coordinate and momentum and since the functions are 0-forms in
differential geometry, we will go one step further in the
exploration of common characteristics of classical and quantum
mechanics, {\it i.e.}, we will find other basic entities of
differential geometry in case of quantum mechanics, see Strocchi
(1966); Heslot (1985) and Ashtekar and Schiling (1997) for related
considerations. We will define n-forms, the Hamiltonian vector
fields and symplectic structure of quantum mechanics and these in
such operatorial forms that they become equivalent to the
corresponding ones of classical mechanics in the semiclassical
limit. This will lead to the conclusion that, regarding all
aspects, classical and quantum mechanics are just two cases of one
general theory. All structural characteristics of these mechanics
will appear to be the same while the crucial difference will be in
operators representing coordinate and momentum.

In the third article we shall propose the symmetrized Poisson bracket
for the case of more than one degree of freedom. This generalization is
not so trivial problem, as it might look like on the first sight,
since it should be related to the dynamical equation for the hybrid
systems - classical and quantum systems in interaction, see Prvanovi\'c
and Mari\'c (2000) and references therein.

Regarding the interaction between classical and quantum systems, it
should be said that one concrete example - the process of
measurement, we have considered in quoted article. The results of this
paper were that the dynamical equation of hybrid systems can produce
noncausal evolution and that the state of quantum system collapses
because of the non-negativity of probability. For finding this the
operator formulation of classical mechanics was extremely useful.
Moreover, this formulation was necessary for the analysis of subsequent
problem: if the equation of motion can produce noncausal evolution,
then what within this equation one can find as responsible for some
nonlinearity which will allowe one to design a formal model of the
collapse. Due to these, the quoted article is strongly connected (or
belongs) to the series of papers concerned with the theory of hybrid
systems being oriented to its application.

\section{Basic definitions}

The present considerations will address a part of kinematical
aspect of quantum and classical mechanics in order to investigate
the semiclassical limit of quantum mechanics, see Shirokov (1976);
Werner (1995) and references therein. For this intention, it is
needed to use the same mathematical framework for both classical
and quantum mechanics. In contrast to the usual approach, where
functions over the phase space were used for representations of
both mechanics and discussions of their relationship, our proposal
is based on the operatorial representations. That is, instead of
transforming the standard operatorial formulation of quantum
mechanics into the formulation comparable to the standard one of
classical mechanics, we expose the most important features of
classical mechanics in the form of operators acting in the same
space where quantum representatives do. The operator formulation
of classical mechanics used here is similar to those given in
Sherry and Sudarshan (1978) and (1979) and Gautam {\it et al.}
(1979). Some connections between our proposal and those given in
Cohn (1980) and (1983); Muga and Snider (1992) and Sala and Muga
(1994) can be found, too.

Our intention is to find such formulations of classical and quantum
mechanics in which the main characteristics of these theories are
preserved. Then, for the intended formulations the following should hold:
1.) the observables and states are in the 1-1 correspondence with the
adequate ones of the standard formulations, 2.) the commutation
relations among observables and the relations among the eigenstates of
observables are the same as are the corresponding ones of the standard
formulations and 3.) the mean values are equal to the corresponding
mean values calculated in standard fashions. The mathematical framework
that will be used can be seen as a direct product  of coordinate and
momentum representations of quantum mechanics. In this way, it will
mimic the phase space of classical mechanics in the most trivial case.

Let us, firstly, define the generalized $h$-dependent operator
algebra as the algebra of polynomials with real coefficients in
operators $\tilde q$, $\tilde p$ and $\tilde I$, which are:
\begin{equation}
\tilde q = \hat q
\otimes \hat I \otimes \left[\hat R_q + \left(1-{h\over
h_o}\right)\hat R_p\right] + \hat I \otimes \hat q \otimes \hat
R_p ,
\end{equation}
\begin{equation}
\tilde p  = \hat p  \otimes \hat I \otimes \hat R_q +
\hat I \otimes \hat p  \otimes \left[\left(1-{h\over h_o}\right)
\hat R_q + \hat R_p \right] ,
\end{equation}
and $\tilde I =\hat I \otimes \hat I \otimes \hat I$. These operators
act in ${\cal H}_q \otimes {\cal H}_p \otimes {\cal H}_r$ with $q$,
$p$ and $r$ being here just the indices of these spaces. The first two
spaces are the rigged Hilbert spaces and the third is a two-dimensional
Hilbert space. More concretely, ${\cal H}_q$ and ${\cal H}_p$ are
formally identical to the rigged Hilbert space of states which is used
in the nonrelativistic quantum mechanics for a single system with one
degree of freedom when the spin is neglected. The indices $q$ and $p$
serve only to denote that the choice of a basis in these spaces is
{\it a priori} fixed when the semiclassical limit is under
considerations. For the basis in ${\cal H}_q \otimes {\cal H}_p$  we
take $\vert q \rangle \otimes \vert p \rangle$. Here, $\vert q \rangle$
and $\vert p \rangle$ are the eigenvectors of $\hat q$ and $\hat p $,
respectively. Then, ${\cal H}_q \otimes {\cal H}_p $ can be seen as an
imitation of the phase space. The third space is introduced only for
the formal reasons and need not to be related to the space of states of
the inner degrees of freedom.

The parameter $h$ takes values from $0$ to $h_o$, $0\le h \le
h_o$, where $h_o$ is attributed to quantum mechanics - the
nonvanishing Planck constant, while, for $h=0$, the above algebra
will be related to the classical mechanics.

The operators $\hat q$ and $\hat p$ are as the operators representing
coordinate and momentum in the standard quantum mechanics. Instead of
reviewing their properties, which can be found in all textbooks of
quantum mechanics, we only mention that they do not commute: $[\hat
q,\hat p ]=i\hbar\hat I$ and that they are the Hermitian (self-adjoint).
For the projectors $\hat R_q$ and $\hat R_p$, the following holds:
$$
\hat R_q \cdot \hat R_p =0, \ \ \hat R_q \cdot \hat R_q =\hat R_q , \ \
\hat R_p \cdot \hat R_p = \hat R_p ,
$$
$$
\hat R_q ^\dagger =\hat R_q , \ \ \hat R_p ^ \dagger =\hat R_p , \ \
\hat R_q +\hat R_p =\hat I ,
$$
$$
\hat R _q =\vert r_q \rangle \langle r_q \vert , \ \ \hat R _p =
\vert r_p \rangle \langle r_p \vert .
$$
They, being introduced to ensure desired properties of the polynomials
in $\tilde q$ and $\tilde p $ for the extreme values of $h$, need not
to have physical meaning.

\section{Quantum mechanics}

When the above algebra of operators is represented with respect to the
basis $\vert q \rangle \otimes \vert p \rangle \otimes \vert r_i
\rangle$, where $i\in \{ q,p\}$ and $\vert r_i \rangle$ is the eigenvector
of $\hat R_i$ for the eigenvalue 1, ($\langle r_i \vert r_j \rangle =
\delta _{i,j}$), then, for $h=h_o$, it becomes equivalent to the
(same representation of) algebra formed over:
\begin{equation}
\hat q_{qm} = \hat q \otimes \hat I \otimes \hat R_q + \hat I \otimes
\hat q \otimes \hat R_p ,
\end{equation}
\begin{equation}
\hat p _{qm} = \hat p  \otimes \hat I \otimes \hat R_q + \hat I \otimes
\hat p  \otimes \hat R_p .
\end{equation}
(We are not going to write these representations explicitly just
for the sake of simplicity of expressions. But, this can be easily
done knowing that $\langle q \vert \hat q \vert q' \rangle = q
\delta (q-q')$, $\langle q \vert \hat p \vert q' \rangle = -i\hbar
{\partial \delta (q-q') \over \partial q}$, $\langle p \vert \hat
p \vert p' \rangle = p \delta (p-p')$ and $\langle p \vert \hat q
\vert p' \rangle = i\hbar {\partial \delta (p-p') \over \partial
p}$.)

The algebra of polynomials with real coefficients in $\hat q
_{qm}$ and $\hat p _{qm}$ and the appropriate eigenvectors are in
the one-to-one correspondence with the algebra of observables and
eigenstates defined in the standard manner in a single rigged
Hilbert space. For example, it holds: $[\hat q_{qm},\hat p
_{qm}]=i\hbar\tilde I$, as it is necessary. On the other hand, due
to the fact that $\hat R_q$ and $\hat R_p$ are idempotent and
mutually orthogonal, the standard representation of an observable,
{\sl e.g.}, $H(\hat q,\hat p )$, is now translated to: $$ H(\hat
q_{qm},\hat p _{qm})=H(\hat q,\hat p )\otimes\hat I \otimes\hat
R_q + \hat I \otimes H(\hat q,\hat p )\otimes\hat R_p. $$ It could
be said that this formulation of quantum mechanics simply doubles
the standard one. Or, it takes the coordinate representation,
multiply it directly with one projector and adds this to the
momentum representation, which was directly multiplied with the
other projector.

If $\vert\Psi_i\rangle$ was the eigenstate of $H(\hat q,\hat p )$
for the eigenvalue $E_i$, then:
$$
\vert \tilde \Psi_i\rangle=c_q\vert\Psi_i\rangle\otimes\vert a\rangle
\otimes\vert r_q\rangle+
c_p\vert b\rangle\otimes\vert\Psi_i\rangle\otimes\vert r_p\rangle,
$$
is the eigenstate of $H(\hat q_{qm},\hat p _{qm})$ for the same
eigenvalue if the coefficients $c_q$ and $c_p$ satisfy the condition
$\vert c_q\vert^2+\vert c_p\vert^2=1$ and if the vectors $\vert
a\rangle$ and $\vert b\rangle$, that are fixed at the beginning of all
considerations being arbitrarily picked, are normalized to one
($\langle a \vert a \rangle = \langle b \vert b \rangle = 1$).
Moreover, it could be easily checked that all relations among
eigenstates ,{\it a la} $\vert \tilde \Psi _i \rangle$, of the same or
different observables are as they were for the corresponding ones, {\it
i.e.}, $\vert \Psi _i \rangle$, of the standard formulation.

Since our intention is to discuss the semiclassical limit of
quantum mechanics after the introduction of new operatorial
representations of both classical and quantum mechanics, it is
necessary to take ${\cal H}_q \otimes {\cal H}_p \otimes{\cal
H}_r$. Of course, this space is much bigger than ${\cal H}$ where
quantum mechanical coordinate and momentum are irreducibly
represented. Only a subspace of ${\cal H}_q \otimes {\cal H}_p
\otimes{\cal H}_r$ that is formed over the basis
$\vert\tilde\Psi_i\rangle$ is interpretable for quantum mechanics.
It depends on the choice of $\vert a\rangle$, $\vert b\rangle$,
$c_q$ and $c_p$ which, after being initially fixed, give desired
irreducible representation. (As will become obvious, the states of
classical system are embedded in the set of operators that act in
${\cal H}_q \otimes {\cal H}_p \otimes{\cal H}_r$.) In other
words, having formal reasons, we have purposely skipped over the
requirement of irreducible representation of quantum mechanical
coordinate and momentum. That this is reducible representation one
can check by finding one nontrivial operator, {\it e.g.}, $\hat q
\otimes \hat I \otimes \hat R_p$, which commutes with $\hat
q_{qm}$ and $\hat p _{qm}$. But, one can recover the
irreducibility in a formal way and/or extract from ${\cal H}_q
\otimes {\cal H}_p \otimes{\cal H}_r$ the physically meaningful
part. This is very similar to the approach developed in Streater
(1966) and references therein where the request of irreducibility
is treated in more relaxed form than usually it is the case.

\section{Classical mechanics}

In the $\vert q \rangle \otimes \vert p \rangle \otimes \vert r_i
\rangle$ representation, but for $h=0$, the general algebra becomes
equivalent to the (same representation of) algebra formed over:
\begin{equation}
\hat q_{cm} = \hat q \otimes \hat I \otimes \hat I ,
\end{equation}
\begin{equation}
\hat p _{cm} = \hat I \otimes \hat p  \otimes \hat I .
\end{equation}
The algebra of polynomials in $\hat q_{cm}$ and $\hat p _{cm}$ with
real coefficients and the appropriate eigenvectors are in the 1-1
correspondence with the standard formulation of classical mechanics
(defined in the framework of functions over phase space). Namely, this
algebra is manifestly a commutative one. To the $c$-number formulation
of a classical variable, {\sl e.g.}, $H(q,p)$, now corresponds:
$$
H(\hat q_{cm},\hat p _{cm}) = H( \hat q \otimes \hat I , \hat I \otimes
\hat p ) \otimes \hat I .
$$

The vectors $\vert q_o\rangle\otimes\vert p_o\rangle\otimes (c_q\vert
r_q\rangle+c_p\vert r_p\rangle)$ are eigenstates of all classical
observables with the real eigenvalues, {\it e.g.}, $H(q_o,p_o)$ for the
above given observable. These vectors now play the role of points in
the phase space. The pure states $\vert q_o\rangle\otimes\vert
p_o\rangle$, taken in dyadic form, can be expressed via $\hat q$ and
$\hat p$ in the following manner:
$$
\vert q_o\rangle\langle q_o\vert\otimes\vert p_o\rangle\langle p_o\vert
=\int \int \delta(q-q_o)\delta(p-p_o) \vert q \rangle \langle q \vert
\otimes \vert p \rangle \langle p \vert dqdp =
$$
$$
=\delta (\hat q-q_o )\otimes \delta (\hat p -p_o ) .
$$
Then, being guided by this, the classical (noncoherently) mixed states
one can define as:
$$
\rho(\hat q\otimes\hat I,\hat I\otimes\hat p )\otimes
(c_q\vert r_q\rangle+c_p\vert r_p\rangle)
(c^*_q\langle r_q\vert+c^*_p\langle r_p\vert).
$$
All classical states will be the Hermitian, non-negative operators and
normalized to $\delta (0) \cdot \delta (0)$ if for $\rho(q,p)$ it holds
that: $\rho (q,p) \in {\bf R}$,  $\rho (q,p) \geq 0$ and $\int \int
\rho (q,p) dqdp = 1$, as it is in the phase space formulation of
classical mechanics, where $\rho (q,p)$ appears in the
coordinate-momentum representation of $\rho(\hat q\otimes\hat I,\hat
I\otimes\hat p )$.

One does not need to proceed with translation of other
features of the classical mechanics into this framework. Having
basic entities, one can do that straightforwardly.

The mean value of the observable, say $\hat A$, when the system
(classical or quantum) is in the state $\hat \rho$, should be
calculated within the theory of hybrid systems by the Ansatz:
\begin{equation}
\langle \hat A \rangle={ {\rm Tr} (\hat \rho \hat A) \over {\rm Tr}
\hat \rho } .
\end{equation}
The norm $\delta (0) \cdot \delta (0)$ of classical states will be
regularized in this way and the mean values will be equal to those
found within the standard formulations for the appropriate variables,
observables and states.

The above, and the fact that the phase space formulation of classical
mechanics appears through the kernels of the operator formulation in
the $\vert q \rangle \otimes \vert p \rangle \otimes \vert r_i \rangle$
representation, can be used as the proof of equivalence of these two
formulations. There will be a complete correspondence between the
$c$-number and this operatorial formulation if the dynamical equation
is defined as the operatorial version of Liouville equation,
where the partial derivations within the Poisson bracket are with
respect to the operators $\hat q_{cm}$ and $\hat p _{cm}$. For
now, let  the dynamical equation for the above formulation of
quantum mechanics be the Schr\"odinger (von Neumann) equation, as
it is in the standard formulation. However, in the next paper we
shall show that for quantum mechanical observables and statistical
operators the commutator can be substituted by the symmetrized
Poisson bracket (see next section). This will implicate that the
dynamical equation of quantum mechanics can be the operatorial version
of Liouville equation, as well.

\section{Concluding remarks}

Since these papers are devoted to the formalism and not to the
phenomenology, we are not going in thorough discussions regarding
the meaning of limit $h \rightarrow 0$, where $h$ is thought to be
the natural constant. Instead, let us illustrate what we mean by
the semiclassical limit. The Hamilton function
$H(q,p)$ of classical mechanics and the Hamiltonian $H(\hat q , \hat p
)$  of quantum mechanics are represented here by $H(\hat q _{cm} , \hat
p  _{cm}  )$ and $H(\hat q _{qm} , \hat p  _{qm})$, respectively. If
they are addressing the same physical system, say the harmonic
oscillator, the last two operators follow from $H(\tilde q , \tilde p
)$ for $h=0$ and $h=h_o$, respectively. Of course, it is understood that
one should work in $\vert q \rangle \otimes \vert p \rangle \otimes
\vert r_i \rangle$ representation which, as we have mentioned, we have
not proceeded here only for the sake of simplicity of expressions.
This means that the semiclassical limit of quantum mechanics is
established through the generalized operator algebra since, for
the one extreme value of $h$, it expresses properties characteristic
for the quantum mechanics and, for the other extreme value of $h$, it
has classical mechanical ones.

The semiclassical limit holds for each polynomial with real
coefficients in $\tilde q $ and $\tilde p $ no matter how these
operators are ordered. This allowed us not to specify explicitly
what is the algebraic product of quantum mechanics. The ordering
rule, or the symmetrized product, being unrelated to the
semiclassical limit, we shall discuss in details in the second
paper of this series. In that article, we shall propose new way of
looking on the Lie bracket of quantum mechanics, too. Here, let us
just mention that the symmetrized product of $\hat q ^n$ and $\hat
p ^m$, denoted by $\circ$, will be the sum of all different
combinations of involved operators divided by the number of these
combinations: $$ \hat q ^n \circ \hat p ^m = {n!m! \over (n+m)!}
(\hat q ^n \hat p ^m + \cdots + \hat p ^m \hat q ^n ), $$ where,
in the parenthesis, there should be all different combinations of
$n+m$ operators. The product of two monomials, let say $\hat q ^a
\circ \hat p ^b$ and $\hat q ^c \circ \hat p ^d$, will be: $$
{(a+c)!(b+d)! \over (a+b+c+d)!}(\hat q ^{a+c} \hat p ^{b+d} +
\cdots + \hat p ^{b+d} \hat q ^{a+c} ). $$ Then, the Lie bracket
of quantum mechanics will be: $$ \{ f(\hat q , \hat p ), g(\hat q
, \hat p  )\} _{\bf S} = {\partial f(\hat q , \hat p )\over
\partial \hat q } \circ  {\partial g(\hat q , \hat p )\over
\partial \hat p } -  {\partial g(\hat q , \hat p )\over
\partial \hat q }\circ {\partial f(\hat q , \hat p )\over \partial \hat p
},
$$
where, within the polynomials $f(\hat q , \hat p )$ and $g(\hat q , \hat
p  )$, the operators of coordinate and momentum are multiplied
according to the symmetrized product $\circ$.

Only after having all these results, we will be able to address in
complete the problem of dequantization: the semiclassical limit of the
kinematical aspect of quantum mechanics and reexpressed dynamical
equation of quantum mechanics will be unified in a way leading to the
proposition of the semiclassical limit of all quantum mechanical
differential geometric entities. Moreover, having defined the
symmetrized product and Poisson bracket, we will be able to propose the
quantization procedure in the most trivial case. In this way, both
directions of the correspondence principle will be covered. The states
are seen as the secondary in our proposal. The meaningful states are
solutions of the appropriate eigenvalue problems and, because quantum
and classical observables are essentially different, they differ, too.

If one is not interested in the semiclassical limit, then the
operator formulation of classical mechanics can be simplified.
Namely, after noticing that the algebra of variables of classical
system with one degree of freedom is isomorphic to the maximal
Abelian subalgebra of operators related to the quantum system with
two degrees of freedom, one can take $\hat q \otimes \hat I$,
$\hat I \otimes \hat p$ and $\hat I \otimes \hat I$ as the basic
elements of the operatorial formulation of classical mechanical
algebra. This  formulation of classical mechanics can find its
applicability in, for example, the analysis of the problem of
measurement.

Similar proposition one can find in Sherry and Sudarshan (1978) and
(1979) and Gautam {\it et al.} (1979). Without repeating the
detailed analysis of these articles, let us mention that a
regulating procedure which recovers the classicality was found as
necessary there and, for this purpose, the so called principle of
integrity was introduced. This is related to the following
proposition of our approach: only the algebra formed over $\tilde
q$, $\tilde p$ and $\tilde I$ should be considered since only to
its elements the physical meaning can be attributed and this only
for the extreme values of Planck constant. All operators different
from these will never occur if they were not introduced with some
unphysical - artificial, reasons.

The framework of functions over the phase space, that is often
used for simultaneous representation of classical and quantum
mechanics and investigation of the semiclassical limit, we find
unsuitable for the related and well-known problems. We believe
that unpleasant situations, the example of which is to end with
possibly negative (quasi)distributions, are the consequence of the
fact that this mathematical ambient is not wide enough to allow
meaningful formulation of quantum mechanics.

The framework of commutative operators, on the other hand,
certainly is not the minimal one needed for the representation of
only classical mechanics. But, motivations for its introduction do
exist. Some problems definitely ask for a framework wider than the
usually used one. Only with the operatorial formulation of
classical mechanics, the argumentation regarding the mentioned
problem of measurement can be proceeded and completed without
going in conflict with physical meaning, see Prvanovi\'c and Mari\'c
(2000).

The space ${\cal H}_q \otimes {\cal H}_p \otimes {\cal H}_r$ we
find not only suitable, but also necessary for the meaningful
representations of both mechanics and discussion of the
semiclassical limit. So, with respect to this, it could be
qualified as the minimal. However, one should be careful not to
misuse its formal opportunities since, if one would do that, and
only then, one would go out of physics. In short, the comparation
of our approach with the alternative ones can be summarized in the
following way: according to our opinion, it is better to be faced
with the need to restrict the considerations and to do that {\it a
priori} than to rectify it {\it a posteriori}.

\end{document}